\documentclass[12pt]{article}
\usepackage{amsfonts,amsmath,amssymb,graphicx}
\usepackage{algorithm, algorithmic}
\usepackage{caption}
\usepackage{comment}
\usepackage{subcaption}
\usepackage{natbib}
\usepackage{color}
\usepackage[dvipsnames,table,dvipsnames*, svgnames*]{xcolor}
\usepackage{epstopdf}

\newcommand{\ignore}[1]{}

\newcommand{\Exp}{\mathbb{E}}
\newcommand{\Var}{\text{Var}}
\newcommand{\Cov}{\text{Cov}}

\newcommand{\Prob}{\mathbb{P}}

\newcommand{\X}{\mathbf{X}}
\newcommand{\bP}{\mathbf{P}}
\newcommand{\bQ}{\mathbf{Q}}
\newcommand{\bS}{\mathbf{S}}
\newcommand{\bG}{\mathbf{G}}
\newcommand{\bSigma}{\boldsymbol{\Sigma}}
\newcommand{\Sighatpi}{\widehat{\bSigma}_{\hat{\pi}}}
\newcommand{\Sigpi}{\bSigma_{\hat{\pi}}}
\newcommand{\bpi}{\boldsymbol{\pi}}
\newcommand{\bV}{\mathbf{V}}
\newcommand{\bU}{\mathbf{U}}
\newcommand{\bLambda}{\mathbf{\Lambda}}

\newtheorem{theorem}{Theorem}
\newtheorem{lemma}{Lemma}

\newtheorem{remark}{Remark}

\begin{document}
\title{A Model for Paired-Multinomial Data and Its Application to  Analysis of Data on a Taxonomic  Tree}
\author{Pixu Shi and Hongzhe Li}

\date{}

\maketitle

\footnotetext[1]{Pixu Shi is with Department of Biostatistics, University of Wisconsin - Madison; Hongzhe Li is with Department of Biostatistics and Epidemiology, University of Pennsylvania. This research was supported by NIH grants CA127334 and GM097505.}

\begin{abstract}
In human microbiome studies, sequencing reads data are often summarized as counts of bacterial taxa at various taxonomic levels specified by a taxonomic tree. This paper considers  the problem of analyzing two repeated measurements of microbiome data from the same subjects. Such data are often collected to  assess the change of microbial composition after certain treatment, or the difference in microbial compositions across body sites. Existing models for such count data are limited in modeling the covariance structure of the counts and in  handling  paired  multinomial count data. A new probability distribution is proposed  for paired-multinomial count data, which allows flexible covariance structure  and can be used to model repeatedly  measured multivariate count data. Based on this distribution,  a test statistic is developed for testing the difference in compositions based on paired multinomial  count data. The proposed  test can be applied to the count data observed on a taxonomic tree in order to test difference in microbiome compositions  and to identify the subtrees with different subcompositions. Simulation results indicate that proposed test has correct type 1 errors and increased power compared to some commonly used methods.  An analysis of  an upper respiratory tract  microbiome data set is used to illustrate the proposed methods. 
\end{abstract}


\section{Introduction}

The human microbiome  includes all microorganisms in and on the human  body \citep{gill2006metagenomic}. These microbes play important roles in  human metabolism in order to maintain  human health. Dysbiosis of gut microbiome has been shown to be associated with many human diseases such as obesity, diabetes and inflammatory bowel disease  \citep{turnbaugh2006obesity,qin2012metagenome,manichanh2012gut}.  Next generation sequencing technologies make it possible to quantify the relative composition of microbes in high-throughout.  Two high-throughput sequencing  approaches have been used in microbiome studies.  One approach is based on sequencing the 16S ribosomal RNA (rRNA) amplicons, where the resulting reads provide information about the  bacterial taxonomic compositions. Another approach is  based on shotgun metagenomic sequencing, which sequences all the microbial genomes presented in the sample, rather than just one marker gene. Both 16S rRNA and shotgun sequencing approaches provide  bacterial taxonomic composition information  and have been widely applied to human microbiome studies, including the Human Microbiome Project \citep{turnbaugh2007human} and the Metagenomics of the Human Intestinal Tract project \citep{qin2010human}.

Compared to shotgun metagenomics, 16S rRNA sequencing is an amplicon-based approach, which makes the detection of rare taxa easier and requires less starting genomic material than the metagenomic approaches.  One important step in analysis of such 16S amplicon sequencing reads data is to assign them to a taxonomy tree.  
Several computational methods are available for accurate taxonomy assignments, including 
BLAST \citep{blast}, the online
Greengenes \citep{greengene} and RDP \citep{RDP} classifiers, and several tree-based
methods.  \cite{Liu} compared several of these methods and recommended use of Greengenes or RDP classifier. Each taxonomy assignment method produces lineage assignments at the levels of domain, phylum, class, order, family and genus. The final data can be summarized as counts of reads that are assigned to nodes of a known taxonomic tree.  

Given the multivariate nature of the count data measured on the taxonomic tree,  methods for analysis of multivariate count  data  are greatly needed in the microbiome research.  Researchers are interested in testing multivariate hypotheses concerning the effects of treatments or experimental factors on the whole assemblages of bacterial taxa.  These types of analyses are useful for studies aiming at assessing the impact of microbiota on human health and on characterizing the microbial diversity in general. 
Multivariate methods for testing  the differences in bacterial taxa composition between groups of metagenomic samples have been developed.  The commonly used methods include permutation test such as Mantel test \citep{Mantel}, Analysis of Similarity (ANOSIM)  \citep{Clarke}, and distance-based  MANOVA (PERMANOVA) \citep{Anderson:2001}.  
An alternative  test is based on the 
Dirichlet multinomial (DM) distribution  to model the counts of sequence reads from microbiome samples
\citep{LaRosa:PLOS:2012, LiChen}.  However, this family of DM probability models may not be appropriate for microbiome data because, intrinsically, such models impose a negative correlation among every pair of taxa. The microbiome data, however, display both positive and negative correlations \citep{DM1}. Models that allow for flexible covariance structures are therefore needed. 

Many microbiome studies involve collection of 16S amplicon sequencing data over time or over different body sites  in order to assess the dynamics of the microbial communities.  Such studies generate paired-multinomial  count data, where the repeatedly observed microbiomes and the corresponding taxonomic count data are dependent.  Modeling such paired-multinomial count data is the focus of our paper.  To the best of our knowledge, there is no flexible model for such paired-multinomial data.  In this paper,  a probability distribution for paired multinomial  count data, which allows  flexible covariance structure, is introduced. The model can be used to model repeatedly  measured multivariate counts.  Based on this paired-multinomial  distribution,  a test statistic is developed  to test the difference of compositions from paired  multivariate count data.    An application of the test to the analysis of count data observed on a taxonomic tree is developed in order to test difference in paired microbiome compositions and to 
identify the subtrees with differential subcompositions. 

The paper is organized as follows. In Section \ref{DM}, the Dirichlet multinomial model and the test of  compositional equality based on this model  are briefly reviewed. A paired  multinomial (PairMN) model for paired count data is defined.   In Section \ref{sec:testing},  a statistical test of equal composition based on the paired  multinomial model is developed  and  is applied to count data observed on a taxonomic tree to test for overall compositional difference and to identify the subtrees that show different subcompositions.  Results from simulation studies are reported in Section \ref{sec:simulation} and application to an analysis of gut microbiome data is given in Section \ref{sec:real}. A brief discussion is given in Section \ref{sec:discuss}.

\section{Paired  Multinomial Distribution of Paired  Multivariate Count Data}\label{DM}
\subsection{Dirichlet multinomial distribution for multivariate count data and the associated two-sample test}
\begin{figure}[H]
	\centering
	\begin{subfigure}[b]{1\textwidth}
		\centering
		\includegraphics[trim=0cm 4.5cm 0cm 4.5cm, clip=TRUE,width=0.75\linewidth]{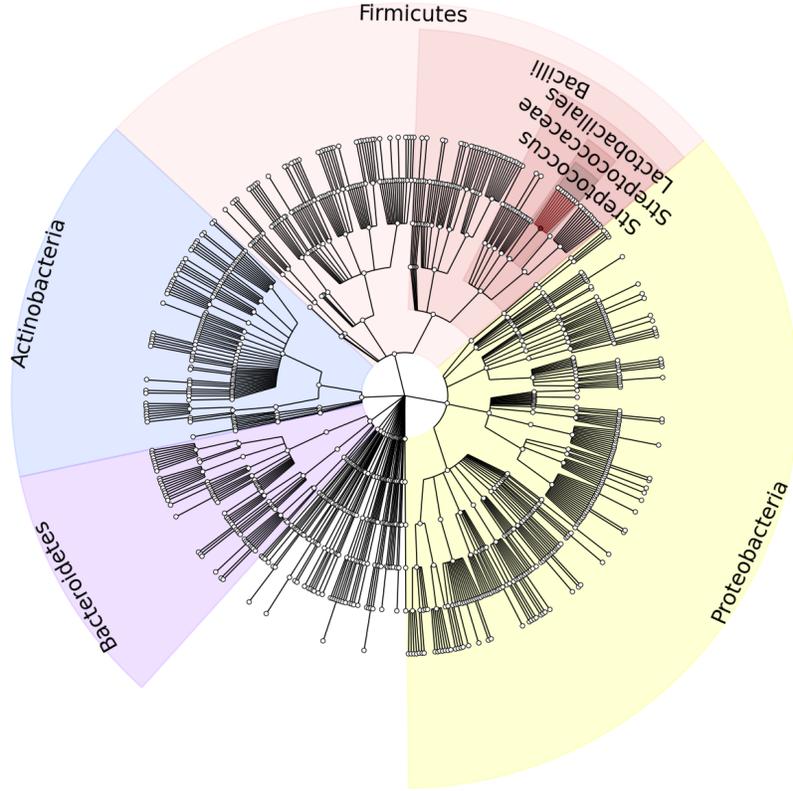}\caption{}
	\end{subfigure}
\vspace{-0.2in}	\begin{subfigure}[b]{1\textwidth}
		\centering
		\includegraphics[width=0.55\linewidth, angle=0]{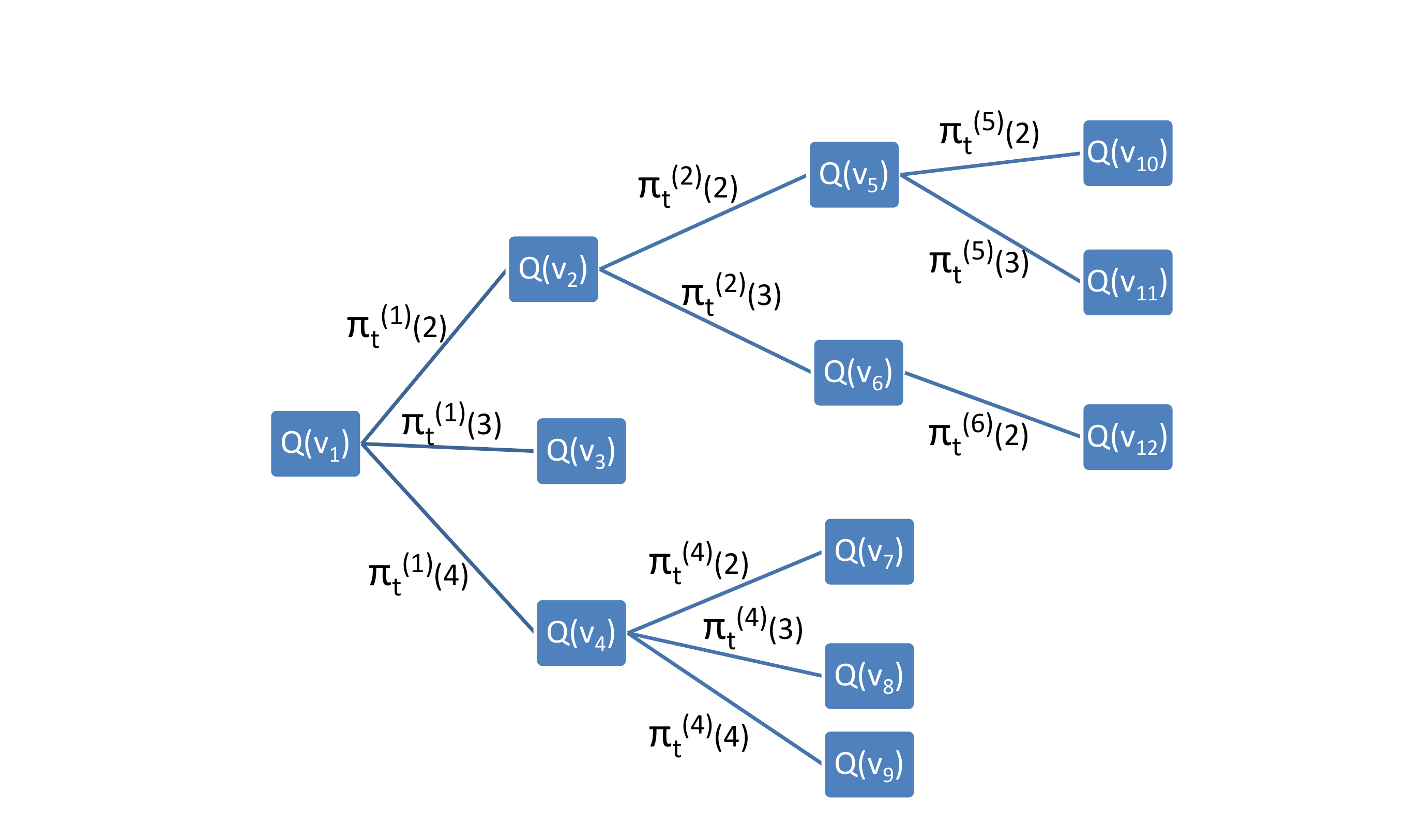}\caption{}
	\end{subfigure}

	\caption{(a) Taxonomic tree in microbiome studies (generated using GraPhlAn \citep{GraphlAn}), where circles from outer to inner are genus, family, order, class, phylum, and kingdom. 
		In our simulations {with sparse differential pattern}, the count of genus {\it Streptococcus} is perturbed to generate the samples from the alternative distribution. As a result, the subtrees with differential subcompositions are the ones with parent node of Kingdom {\it Bacteria},  Phylum {\it Firmicutes},  Class {\it Bacilli},  Order {\it Lactobacillales},  Family {\it Streptococcaceae} and
		 Genus {\it Streptococcus}. 
	(b) An	illustration of the probability model of the counts on taxonomic tree, where 
		$\X_{it}^{(1)}=(\bQ_{it}(v_{2}),\bQ_{it}(v_{3}), \bQ_{it}(v_{4}), \bQ_{it}(v_{1})-\sum_{j=2}^4 \bQ_{it}(v_{j})),
		\X_{it}^{(2)}=(\bQ_{it}(v_{5}),\bQ_{it}(v_{6}), \bQ_{it}(v_{2})-\bQ_{it}(v_{5})-\bQ_{it}(v_{6})),
		N_{it}^{(1)}=\bQ_{it}(v_{1}) = \textbf{1}^\top \X_{it}^{(1)},
		N_{it}^{(2)}=\bQ_{it}(v_{2}) = \textbf{1}^\top \X_{it}^{(2)}.$}
	\label{fig:data_generation}
\end{figure}

Consider a set of microbiome samples measured on  $n$ subjects, where for each sample, the 16S rRNA sequencing reads are aligned to the nodes of an existing taxonomic tree \citep{Liu} (see Figure \ref{fig:data_generation} (a)). Consider a subtree defined by  one internal node of the tree with $d-1$ child nodes.  
Let $\X_1,\dots,\X_n \in \mathbb{N}^{d}$ denote the  $d$-dimensional count data of these $n$ samples, where the $j$th entry of $\X_i$ is the number of the sequencing reads aligned to the $j$th child node from the $i$th sample and the last element  of $\X_i$ is the number of the sequencing reads aligned to the internal node.  The following model is developed assuming that  $d<n$. Section \ref{sec:global} presents further details on how the model and the test proposed in this Section  can be applied to data from the whole taxonomic  tree.

 In order to account for overdispersion of the count data in microbiome studies,  $\X_1,\dots,\X_n$ are often assumed to follow a Dirichlet multinomial distribution \citep{LaRosa:PLOS:2012, LiChen}, $DM(N_i, \boldsymbol{\alpha}, \theta), i=1,\cdots, n$, where $N_i$ is the total number of the reads from the $i$th sample that are mapped to these $d$ taxa,  $\boldsymbol{\alpha}=(\alpha_1,\cdots,\alpha_d)$, $0\le \alpha_j\le 1$, $\sum_j \alpha_j=1$ is a vector of the true  subcomposition of the taxa of a given subtree, and $\theta$ is an overdispersion parameter.

Consider the two-group comparison problem, where the count data of two groups of microbiome samples, denoted by  $\X_{11},\dots,\X_{n_11}$ for the $n_1$ samples in group 1 and $\X_{12},\dots,\X_{n_22}$ for the $n_2$ samples in group 2 are given.  { Assuming $n_1,n_2>d$}, \cite{LaRosa:PLOS:2012} assume each  independently follows a DM distribution with
\begin{equation}\label{eq:DM}
\begin{split}
& \X_{i1}\sim DM(N_{i1},\boldsymbol{\alpha}_1,\theta_1), i=1,\dots, n_1,\\
& \X_{i2}\sim DM(N_{i2},\boldsymbol{\alpha}_2,\theta_2), i=1,\dots, n_2,
\end{split}
\end{equation}
and propose a test for the following hypothesis of equal subcomposition: 
\begin{equation}\label{eq:hypothesesOld}
H_0:\boldsymbol{\alpha}_1=\boldsymbol{\alpha}_2 
\mbox{ vs } H_a: \boldsymbol{\alpha}_1\neq\boldsymbol{\alpha}_2. 
\end{equation}
Define
\begin{equation}\label{eq:pi}
\hat{\bpi}_t = (\sum_{i=1}^{n_t}\X_{it})/(\sum_{i=1}^{n_t}N_{it}), t=1,2,
\end{equation}
which is a consistent estimator for $\boldsymbol{\alpha}_t$ for $t=1,2$.
\cite{Wilson:JRSSC:1989} and \cite{LaRosa:PLOS:2012} proposed to reject the null hypothesis when 
\begin{equation}\label{eq:old_test}
\sum_{k=1}^d\dfrac{(\hat{\pi}_{1k}-\hat{\pi}_{2k})^2}{C_1\hat{\pi}_{1k}+C_2\hat{\pi}_{2k}} > {\chi^2_{d-1}}^{-1}(1-\alpha),
\end{equation}
where
$$C_t=\dfrac{1}{N_{\cdot t}^2}\Big(\hat{\theta}_{\alpha_t}\big(\sum_{i=1}^{n_t} N_{it}^2-N_{\cdot t}\big)+N_{\cdot t}\Big), \quad t=1,2$$
and $\hat{\theta}_{t}$ is a consistent estimator of $\theta_{t}, t=1,2$.

In many microbiome studies,  microbiome data are often observed for the same subjects over two different time points or different body sites. If the microbiome of each subject is measured several times,  these repeated measurements are not independent to each other and cannot be handled by the independent DM model. Thus, a new model is developed in the next section to take into account the within  subject correlations.

\subsection{Paired Multinomial Distribution for Paired Multinomial Data}\label{sec:CorM_model}
Any model for paired multinomial data such as those observed in  microbiome studies with repeated measures needs to account for the dependency of the data. 
For a paired multinomial random variable  $\X_i=(\X_{i1}, \X_{i2})\in\mathbb{N}^{d\times 2}, i=1,\dots, n$,  a paired  multinomial (PairMN) distribution can be defined as 
$$\X_i\sim\mbox{PairMN}\big(N_{i1},N_{i2}, \bpi_1,\bpi_2, \bSigma_1, \bSigma_2, \bSigma_{12}\big),$$
where 
\begin{equation}\label{eq:CorM}
\begin{split}
\X_{it}| \bP_{it}&\sim~ \mbox{Multinomial}(N_{it}, \bP_{it})\in \mathbb{R}^{d},\\
\Exp \bP_{it}&=~\bpi_t,\\
\Var \bP_{it}&=~\bSigma_t,\\
\Cov(\bP_{i1}, \bP_{i2})&=~\bSigma_{12},
\end{split}
\end{equation}
for $t=1,2$. Here, the group-specific subcomposition is represented by $\bpi_t$. 
The joint distribution of $(\bP_{i1},\bP_{i2})$ is only defined up to its first and second moments so that it  includes a wide range of distributions. 

Under this probability model, the moments of $ \X_{it} $ are given as follow: 
\begin{equation}\label{eq:moments}
\begin{split}
\Exp \X_{it} &=~ N_{it}\bpi_t,\\
\Var \X_{it} &=~ N_{it}\big(diag(\bpi_t)-\bpi_t\bpi_t^\top\big)+N_{it}(N_{it}-1)\bSigma_t,\\
\Cov(\X_{i1}, \X_{i2}) &=~ N_{i1}N_{i2}\bSigma_{12}.
\end{split}
\end{equation}

Compared to the DM model in \eqref{eq:DM}, this model has several important features. First, for a given $t$, the model allows a more flexible covariance structure for the observed counts that is characterized by  $\bSigma_t$.  
Second, this model uses $\bSigma_{12}$ to quantify the correlation between the repeated samples of the same subject. 
If $\bP_{it}$ is assumed to follow a Dirichlet distribution, the proposed model in \eqref{eq:CorM} becomes the DM distribution in \eqref{eq:DM}. However,  a parametric assumption is not needed to achieve the flexible covariance structure. 

\section{Statistical Test Based on  Paired Multinomial Samples}\label{sec:testing}
For a given subtree of $d$ taxa, in order to  test if there is any difference in microbiome subcompositions between two correlated samples,   consider the following hypotheses:
\begin{equation}\label{eq:hyptotheses2}
H_0: \bpi_1=\bpi_2 \mbox{ vs } H_a:  \bpi_1 \neq \bpi_2.
\end{equation}
Define $$\hat{\bpi}_t = \dfrac{\sum_{i=1}^n \X_{it}}{\sum_{i=1}^n N_{it}},$$ then
 $\Exp \hat{\bpi}_t = \bpi_t$.    A Hotelling's $T^2$ type of statistic based on $\hat{\bpi}_1 - \hat{\bpi}_2$ can then be developed.  
 
 Assume that the sample size $n$ is greater than the number of bacterial taxa $d$ considered. A consistent estimator for $\Sigpi=\Var\left(\hat{\bpi}_1 - \hat{\bpi}_2\right)$ is given in the following Lemma.
\begin{lemma}\label{th:covariance}
Define
\begin{equation*}
\begin{split}
N_{\cdot t} &= \sum_{i=1}^nN_{it}\\
N_{ct}&=\dfrac{1}{(n-1)N_{\cdot t}}\left(N_{\cdot t}^2-\sum_{i=1}^n N_{it}^2\right)\\
\bS_t&=\dfrac{1}{n-1}\sum_{i=1}^n N_{it}(\hat{\bpi}_{it}-\hat{\bpi}_t)(\hat{\bpi}_{it}-\hat{\bpi}_t)^\top\\
\bG_t&=\dfrac{1}{N_{\cdot t}-n}\sum_{i=1}^n N_{it}\big(diag(\hat{\bpi}_{it})-\hat{\bpi}_{it}\hat{\bpi}_{it}^\top\big)\\
\widehat{\bSigma}_{12}&=\dfrac{1}{(n-1)}\sum_{i=1}^n\dfrac{N_{i1}+N_{i2}}{N_{c1}+N_{c2}}(\hat{\bpi}_{i1} - \hat{\bpi}_{1})(\hat{\bpi}_{i2}-\hat{\bpi}_{2})^T
\end{split}
\end{equation*}
where $\hat{\bpi}_{it} = \X_{it}/N_{it}$. { Assuming $N_{it}$'s are bounded by a common fixed number $N$ for all $i$ and $t$,} then
\begin{equation}\label{eq:sigmahat}
\begin{split}
\Sighatpi =& \sum_{t=1}^2\left\{\dfrac{\bS_t+(N_{ct}-1)\bG_t}{N_{ct}N_{\cdot t}}+\dfrac{\sum_{i=1}^nN_{it}^2-N_{\cdot t}}{N_{ct}N_{\cdot t}^2}(\bS_t-\bG_t)\right\}\\
&-\dfrac{\sum_{i=1}^n N_{i1}N_{i2}}{N_{\cdot 1}N_{\cdot 2}}\big(\widehat{\bSigma}_{12}+\widehat{\bSigma}_{12}^\top\big)
\end{split}
\end{equation}
is a consistent estimator of $\Sigpi =\Var\left(\hat{\bpi}_1 - \hat{\bpi}_2\right)$. In other words,
\begin{equation}
||\Sighatpi-\Sigpi||_{\max}\rightarrow 0 \mbox{ in probability as }n\rightarrow\infty
\end{equation}
where $||\cdot||_{\max}$ is the max norm of a matrix.
\end{lemma}

Since $\Sighatpi$ is singular due to the unit sum constraint on $\bP_{it}$, 
 a statistic to test $H_0$ vs $H_a$  specified in \eqref{eq:hyptotheses2} is defined as 
\begin{eqnarray}\label{eq:teststat}
F = \dfrac{n-d+1}{(n-1)(d-1)}(\hat{\bpi}_1 - \hat{\bpi}_2)\Sighatpi^\dagger(\hat{\bpi}_1 - \hat{\bpi}_2)^\top,
\end{eqnarray}
where $\Sighatpi^\dagger$ is the Moore-Penrose pseudoinverse of $\Sighatpi$. The Moore-Penrose pseudoinverse is chosen over other forms of pseudoinverse because of its simple expression related to the singular values of the original matrix and inverted matrix.  Since  $\Sighatpi$ is not guaranteed to be non-negative definite,  the negative eigenvalues of $\Sighatpi$ are truncated to 0 in the computation, however, the truncation does not affect the convergence of $\Sighatpi$.

The following theorem shows that under the null, the test statistic defined in \eqref{eq:teststat} follows an asymptotic $F$-distribution with degrees of freedom of $d-1$ and $n-d-1$.
\begin{theorem}\label{th:test}
With test statistic $F$ defined in \eqref{eq:teststat}, an asymptotic level $\alpha$ test for testing \eqref{eq:hyptotheses2} is to reject $H_0$ when
\begin{equation}\label{eq:test}
F>F_{d-1,n-d+1}^{-1}(1-\alpha).
\end{equation}
The $p$-value for testing \eqref{eq:hyptotheses2} is
\begin{equation}\label{eq:pval}
p = 1 - F_{d-1,n-d+1}(F).
\end{equation}
\end{theorem}

\begin{remark}
Lemma~\ref{th:covariance} and the proposed test statistic in \eqref{eq:teststat} can be easily extended to unpaired multivariate count data with unequal sample sizes. {Specifically,    $\Sighatpi$ in \eqref{eq:sigmahat} can be replaced by 
$$\Sighatpi = \sum_{t=1}^2\left\{\dfrac{\bS_t+(N_{ct}-1)\bG_t}{N_{ct}N_{\cdot t}}+\dfrac{\sum_{i=1}^nN_{it}^2-N_{\cdot t}}{N_{ct}N_{\cdot t}^2}(\bS_t-\bG_t)\right\}$$
and  $\bS_t$, $\bG_t$, $N_{\cdot t}$ and $N_{ct}$ within each group $t$ can be calculated in the same fashion as Lemma~\ref{th:covariance}.
}
\end{remark}

\section{Analysis of Microbiome Count Data Measured on the Taxonomic Tree}\label{sec:global}

This section presents details of applying the proposed F-test in Theorem \ref{th:test} for paired-multinomial data to  analysis of 16 S data. Our goal is  to identify the subtrees of a given taxonomic tree that show differential  subcomposition between two repeated  measurements and to perform a global test of overall microbiome composition between two conditions. A global probability model for count data on a taxonomic tree is first introduced. 

\subsection{A global probability model for count data on a taxonomic tree}\label{sec:global_model}
A rooted taxonomic tree $T$ with nodes $v_1,\dots, v_{K_0}$ representing for the taxonomic units of $T$ is often available based on 16S sequencing data.  For   each microbiome sample, the 16S  reads can be aligned to the nodes of $T$ to output the count of reads assigned to each node. 
Without loss of generality, assume that the first $K$ nodes $v_1,\dots,v_K$ are  all the internal non-leaf nodes and $v_1$ is  the root node. Also, denote $\tau(v_k)$ as the set of all direct child nodes of $v_k, k=1,\dots,K$.  Figure ~\ref{fig:data_generation} (b)  presents a tree to illustrate  the setup. 

For a given internal node $v_k$, let $\bQ(v_k)$ be the sum of number of reads assigned to $v_k$ and the number of reads assigned to  all its descending nodes.  {For example, if $v_k$  corresponds to the phylum {\it  Firmicutes},  $\bQ(v_k)$ is  the count of all reads  assigned to {\it Firmicutes} and all classes that belong to {\it Firmicutes}. For convenience, denote $\bQ(S)=\big(\bQ(v_{k_1}), \dots, \bQ(v_{k_j})\big)$ for any set of nodes $S=\{v_{k_1},\dots, v_{k_j}\}$.
For each split from a parental node to the child nodes, the reads on the parent node are either assigned to a child node or remain unassigned. 
For each parent node $v_k$,  let the vector $\bQ(\tau(v_k))$ denote  the counts of reads assigned to its direct child nodes and $\bQ(v_k)-\sum_{j\in\tau(v_k)}\bQ(v_j)$ be the count of reads that can only be assigned to $v_k$. 
For a subject $i$ with measurement index $t$, at  a given  internal node $v_k$, $k=1,\dots,K$, denote   
\begin{equation}
\begin{split}
\X_{it}^{(k)}&=\big(\bQ_{it}(\tau(v_k)), \bQ_{it}(v_k)-\sum_{j:v_j\in\tau(v_k)}\bQ_{it}(v_j)\big)^{\top},\\
N_{it}^{(k)}&=\bQ_{it}(v_k) = \textbf{1}^\top\X_{it}^{(k)},
\end{split}
\end{equation}
{
where   $N_{it}^{(k)}$ is the sum of these read counts.   }

For a  repeated microbiome study, $\bQ_{i1}(T)$ and $\bQ_{i2}(T)$ represent the counts assigned to the nodes of the tree $T$. These count data are assumed to be generated hierarchically, conditioning on the total read count of each internal node.    At each internal node $v_k$,  $k=1,\dots,K$, given the total counts $(\bQ_{i1}(v_k), \bQ_{i2}(v_k))=(N_{i1}^{(k)}, N_{i2}^{(k)})$,  the paired vectors of read counts 
\begin{gather*}
(\X_{i1}^{(k)}, \X_{i2}^{(k)})=\\
\Big(\big(\bQ_{i1}(\tau(v_k)), \bQ_{i1}(v_k)-\sum_{j:v_j\in\tau(v_k)}\bQ_{i1}(v_j)\big)^{\top}, \big(\bQ_{i2}(\tau(v_k)), \bQ_{i2}(v_k)-\sum_{j:v_j\in\tau(v_k)}\bQ_{i2}(v_j)\big)^{\top}\Big)
\end{gather*}
 is assumed to  follow a PairMN distribution 
\begin{equation}
\begin{split}
(\X_{i1}^{(k)}, \X_{i2}^{(k)})|(N_{i1}^{(k)}, N_{i2}^{(k)})\sim \mbox{PairMN}\big(N_{i1}^{(k)}, N_{i2}^{(k)}, \bpi^{(k)}_1,\bpi^{(k)}_2, \bSigma^{(k)}_1, \bSigma^{(k)}_2, \bSigma_{12}^{(k)}\big).
\end{split}
\end{equation}

As an illustration, for the tree  in Figure~\ref{fig:data_generation} (b),  
the parameters associated with  subtree under node $v_1$ are $\bpi_t^{(1)}=\Exp\left[\X_{it}^{(1)}/N_{it}^{(1)}\big|N_{it}^{(1)}\right]$, which represent the subcomposition  of nodes $v_2$ to $v_4$ and  taxa can not be further assigned.   Similarly $\bpi_t^{(2)}=\Exp\left[\X_{it}^{(2)}/N_{it}^{(2)}\big|N_{it}^{(2)}\right]$ characterizes the subcomposition  of nodes $v_5$ and $v_6$ under the subtree of node  $v_2$.

\subsection{Identification of subtrees of with differential subcompositions based on the proposed test}\label{sec:sub_tree}
In order to identify the subtrees with differential subcompositions  between the two measurements,  the following hypotheses are tested using the  F-test in Theorem~\ref{th:test},
\begin{equation}\label{eq:testing_single}
H_0^{(k)}: \bpi_1^{(k)} = \bpi_2^{(k)}, \quad k=1,\dots,K. 
\end{equation}

Define $p_k$ as the $p$-value from testing $H_0^{(k)}$.  Theorem~\ref{th:test} shows that under the null hypotheses, $p_k$'s are asymptotically uniformly distributed. In fact, they are also asymptotically independent under the null. Take Figure~\ref{fig:data_generation} (b) as an example, under the $H_0^{(1)}$ and $H_0^{(2)}$, 
\begin{equation*}
\begin{split}
\Prob(p_{1}\leq \alpha, p_{2}\leq \beta)&=\int\Prob(p_{1}\leq \alpha|\bQ(v_{2}), p_{2}\leq \beta)\Prob(p_{2}\leq \beta|\bQ(v_{2}))dF(\bQ(v_{2}))\\
&=\int\Prob(p_{1}\leq \alpha|\bQ(v_{2}))\Prob(p_{2}\leq \beta|\bQ(v_{2}))dF(\bQ(v_{2}))\\
&\stackrel{a}{=}\beta\int\Prob(p_{1}\leq \alpha|\bQ(v_{2}))dF(\bQ(v_{2}))\\
&=\beta\Prob(p_{1}\leq \alpha)\stackrel{a}{=}\Prob(p_{1}\leq \alpha)\times\Prob(p_{2}\leq \beta),
\end{split}
\end{equation*}
where $\stackrel{a}{=}$ represents  equation that holds asymptotically.
Therefore, to control for multiple comparisons, the false discovery rate (FDR) procedure \citep{Benjamini:1995} can be used to identify the subtrees with different subcompositions between two repeated measurements.

\subsection{Global test for differential overall  compositions  on taxonomic tree}\label{sec:testing_tree}
The goal for testing the global difference in taxonomic composition between a pair of measurements can be formulated as  the following composite hypothesis, 
\begin{equation}\label{eq:treehypotheses}
H_0: \bpi_1^{(k)} = \bpi_2^{(k)}, k=1,\cdots,K \quad \mbox{vs}\quad H_a: \bpi_1^{(k)} \ne \bpi_2^{(k)}, \mbox{for at least one $k$}. 
\end{equation}
As shown in previous section, under the $H_0$, $p$-values for testing $H_0^{(k)}$ for $k=1,\cdots,K$ are independent.  In addition, the number of tests $K$ is determined by the prior taxonomic tree and does not depend on the sample size $n$. To test this composite hypothesis \eqref{eq:treehypotheses},  a combined $p$-value can be obtained using the Fisher's method,
\begin{equation}\label{eq:combined_fisher}
p_{combined} = 1 - (\chi^2_{2K})^{-1}\left(-2\sum_{k=1}^K\log p_k\right).
\end{equation}
Alternatively, let $p_{(2)}$ be the $2^{nd}$ smallest $p$-value of $p_1,\dots,p_K$, a statistic based on this  $2^{nd}$ smallest $p$-value,
\begin{equation}\label{eq:combined_2nd}
p_{combined} = 1 - \big[1+(K-1)p_{(2)}\big](1-p_{(2)})^{K-1}
\end{equation}
can also be used, where 
{\eqref{eq:combined_2nd} is a special case of Wilkinson's method of $p$-value combination \citep{wilkinson1951statistical} summarized by \cite{zaykin2002truncated}.}
Under the null, the $p_{combined}$ computed using either method is asymptotically uniformly distributed. Test \eqref{eq:combined_2nd} is more powerful if only a small number of subtrees show differential subcomposition   between the two measurements, while test \eqref{eq:combined_fisher} is more suitable if the differences occur in a large number of subtrees.

\section{Simulation Studies}\label{sec:simulation}
\subsection{Comparison with test based on the DM model}\label{sec:simu_single}
To compare the performance of our pairMN test statistic in \eqref{eq:test} with the original unpaired statistic \eqref{eq:old_test}, two data generating models within the class of PairMN are considered. 
The first model generates $\bP_{it}, i=1,\dots, n$ based on a mixture of Dirichlet distributions:
\begin{equation}\label{eq:CorM_Dir}
\begin{split}
& \bP_{it}=(1-\rho)\bP_{it}' + \rho \bP_{i}'', \quad t=1,2,\\
& \bP_{it}'\sim\mbox{Dir}(\boldsymbol{\alpha}_t, \theta_{\alpha_t}), \quad t=1,2,\\
& \bP_{i}''\sim\mbox{Dir}(\boldsymbol{\ell}, \theta_{\ell}).
\end{split}
\end{equation}
Under this setting, 
\begin{equation}
\begin{split}
\bpi_t&=~(1-\rho)\boldsymbol{\alpha}_t + \rho\boldsymbol{\ell}, \quad 0<\rho<1,\quad t=1,2\\
\bSigma_t&=~(1-\rho)^2\theta_{\alpha_t}\big(diag(\boldsymbol{\alpha}_t)-\boldsymbol{\alpha}_t\boldsymbol{\alpha}_t^\top\big) + \rho^2\theta_{\ell}\big(diag(\boldsymbol{\ell})-\boldsymbol{\ell}\boldsymbol{\ell}^\top\big), \quad t=1,2\\
\bSigma_{12}&=~\rho^2\theta_{\ell}\big(diag(\boldsymbol{\ell})-\boldsymbol{\ell}\boldsymbol{\ell}^\top\big).\\
\end{split}
\end{equation}
In our simulation,  the dimension is set as $d=8$.
The parameter  $\rho$ is used to control the degree of correlation in $\bSigma_{12}$, where $\rho$ ranges from 0 to 0.6.  Other parameters are set as  $\theta_{\ell}=1$, $\theta_{\alpha_1}=3$, $\theta_{\alpha_2}=5$, $\boldsymbol{\ell}=(0.12, 0.06, 0.08, 0.43, 0.02, 0.14, 0.1, 0.05)$,  $\boldsymbol{\alpha}_1$ and $\boldsymbol{\alpha}_2$ such that $\bpi_1=(0.15, 0.05, 0.22, 0.3, 0.03, 0.1, 0.07, 0.08)$, and under the alternative hypothesis $\bpi_2=(0.1, 0.1, 0.22, 0.3, 0.03, 0.1, 0.07, 0.08)$. The number of total counts $N_{it}$ are simulated from a Poisson distribution with a mean 1000.  When $\rho=0$, this model degenerates to the Dirichlet-multinomial distribution.

The second model generates $\bP_{it}, i=1,\dots,n$ based on a log-normal distribution. Specifically,
\begin{equation}\label{eq:CorM_Norm}
\bP_{it} =\dfrac{e^{\mathbf{Z}_{it}}}{\boldsymbol{1}^\top e^{\mathbf{Z}_{it}}}, \quad t=1,2,
\end{equation}
where
\begin{equation*}
\begin{split}
(Z_{i1j}, Z_{i2j}) &\sim~ N\left(\left[\begin{array}{c}
\mu_{j1} \\ 
\mu_{j2}
\end{array}\right] , \left[\begin{array}{cc}
\sigma_{j1}^2 & \rho\sigma_{j1}\sigma_{j2}\\ 
\rho\sigma_{j1}\sigma_{j2} & \sigma_{j2}^2
\end{array}\right] \right), \quad j=1,\dots,d,\\
\mathbf{Z}_{it} &=~ (Z_{it1}, \dots, Z_{itd})^\top, \quad t=1,2.
\end{split}
\end{equation*}
Under this setting, no explicit expressions for $\bpi_t$, $\bSigma_t$ and $\bSigma_{12}$ are available, but the correlation can be quantified using $\rho$, and the difference in $\bpi_t$ can be quantified by the difference in $\boldsymbol{\mu}_t=(\mu_{1t},\dots,\mu_{dt})^\top, t=1,2$.
In our simulation,  the  dimension of sample is $d=8$, $\rho$ ranges from 0 to 0.6, $\boldsymbol{\sigma}_t=(\sigma_{1t},\dots,\sigma_{dt})^\top=(1,\dots,1)$ for $t=1,2$, $\boldsymbol{\mu}_1=(3,1,0.5,1,0,1,1,0)$, and $\boldsymbol{\mu}_2=(3,1,1,0.5,0,1,1,0)$ under the alternative. The number of total counts $N_{it}$ are also simulated from a Poisson distribution with a mean 1000.

For both data generating models, sample sizes of $n=20, 50$ and  $100$ are considered. { The dimension of parameters is chosen to be eight in all simulations to mimic the fact that most of the nodes on the taxonomic tree in our study have less than ten child nodes.} The simulations are  repeated  5,000 times for each specific setting and the null hypothesis is rejected at level of $\alpha=0.05$. The type I error and the empirical power of the various tests are shown in Figure~\ref{fig:size_power_single}. 
 It shows that both tests have test size under the nominal level in all  settings. For data simulated from the paired  multinomial-Dirichlet distribution  \eqref{eq:CorM_Dir}, the power of the unpaired test is slightly better than the paired test only when $\rho$ is very small, that is, when there is a  weak  within-subject correlation  (Figure \ref{fig:size_power_single} (a)). This is expected  since the unpaired test \eqref{eq:old_test} is developed  specifically for the Dirichlet-multinomial distribution, i.e. PairMN model with $\rho=0$. When $\rho$ increases from 0 to 0.6, the paired test has a steadily increasing power with the test size still around the nominal level, while the size and power of the unpaired test gradually decrease. The results suggest that compared with the paired test, the unpaired test tends to be conservative and therefore has reduced power in detecting the difference in compositions when the within-subject correlation is large.

For data simulated from log-normal-based PairMN model \eqref{eq:CorM_Norm},  the power of our paired test is much larger than the power of the unpaired test for all values of $\rho$, while the type 1 errors are well controlled (Figure~\ref{fig:size_power_single} (b)). 
These results show that the proposed paired test performs well in both data generating models, suggesting  that our test is very flexible and robust  to  different distributions of $\bP_{it}$.

\begin{figure}[H]
	\begin{subfigure}[b]{1\textwidth}
		\centering
		\includegraphics[width=1\linewidth]{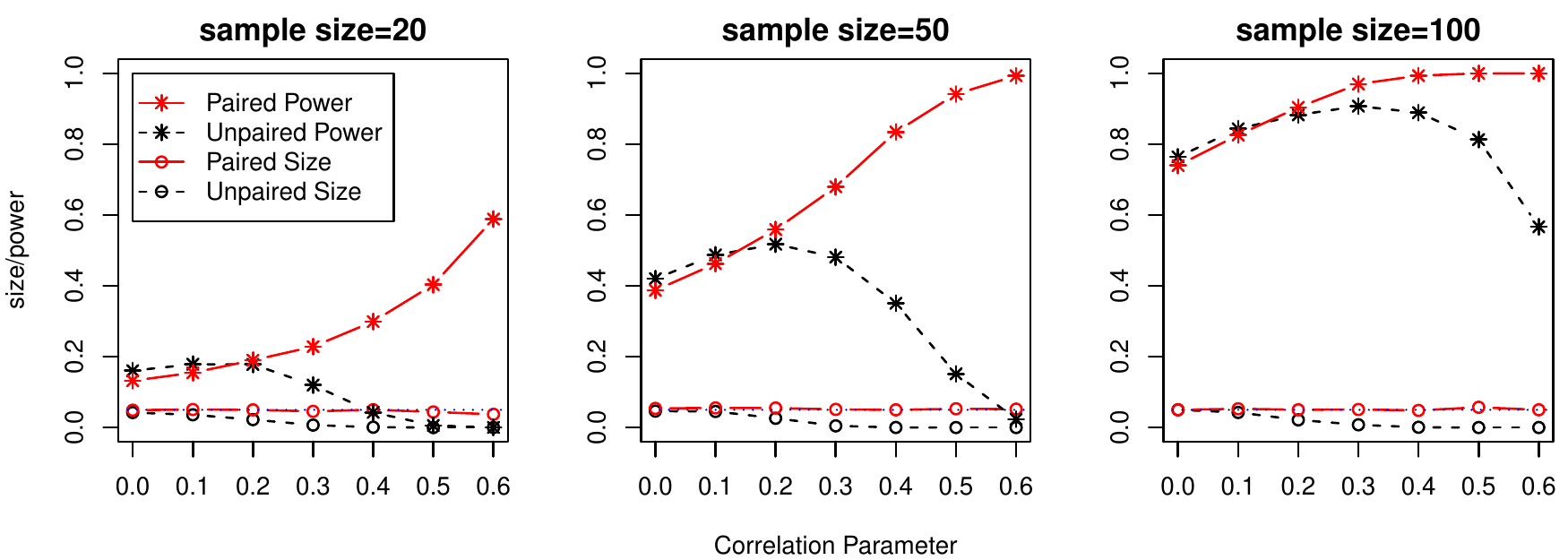}
		\caption{Mixed Dirichlet $P_{it}$.}
		\vspace{0.5cm}
	\end{subfigure}
	\begin{subfigure}[b]{1\textwidth}
		\centering	\includegraphics[width=1\linewidth]{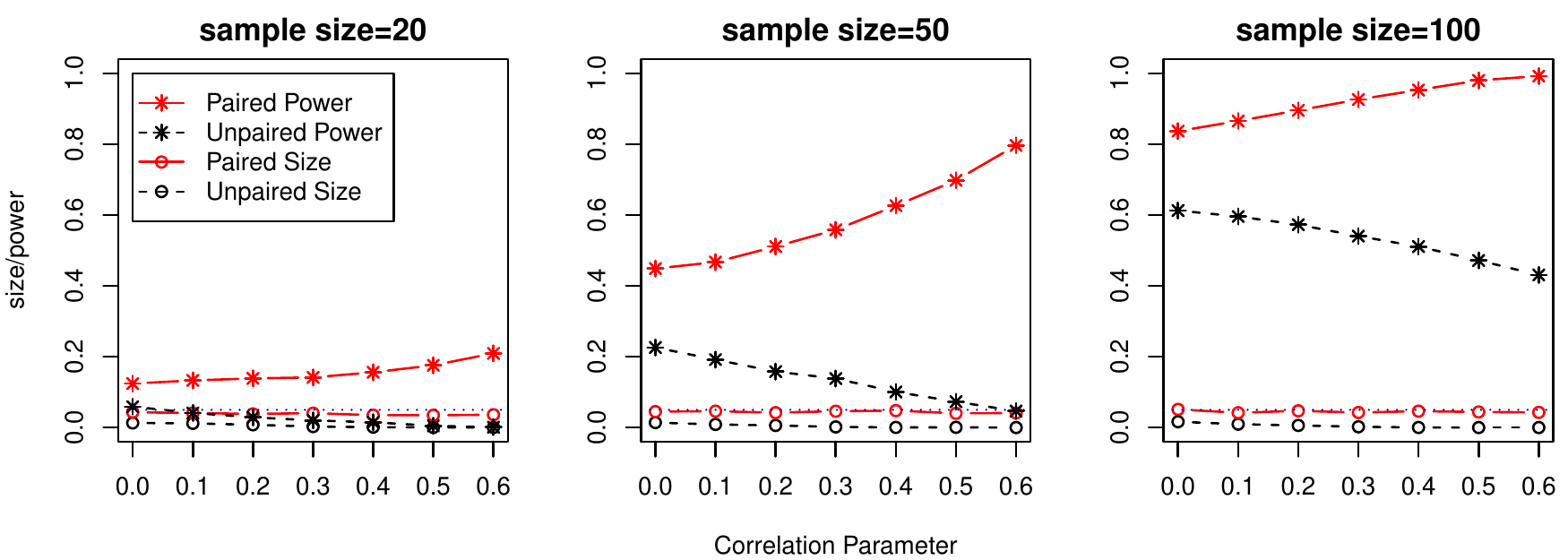}
		\caption{Log-normal based $P_{it}$.}
	\end{subfigure}
	\caption{Simulation results: size and power of the paired and unpaired tests for data simulated under the PairMN model (a) and the correlated log-normal model (b) for sample size $n=20, 50$ and 100.  x-axis is the correlation parameter $\rho$.}
	\label{fig:size_power_single}
\end{figure}

\subsection{Simulating count data on a taxonomic tree }\label{sec:simu_tree}
The proposed  tests in \eqref{eq:combined_fisher} and \eqref{eq:combined_2nd} are further compared with PERMANOVA test \citep{Anderson:2001} using $L^1$ Kantorovich-Rubinstein (K-R) distance \citep{Evans:JRSSB:2012} with unit branch length and with each pair of samples as a stratum. 
Using the notations in Section~\ref{sec:testing_tree}, the $L^1$ K-R distance between two trees $\bQ_{i_1t_1}$ and $\bQ_{i_2t_2}$ is given by 
\begin{equation}\label{eq:KR}
d(\bQ_{i_1t_1}, \bQ_{i_2t_2})=\sum_{k=1}^{K_0}\left|\mathbf{p}_{i_1t_1}(v_k)-\mathbf{p}_{i_2t_2}(v_k)\right|
\end{equation}
where
$$\mathbf{p}_{it}(v_k)=\big(\bQ_{it}(v_k)-\sum_{j:v_j\in\tau(v_k)}\bQ_{it}(v_j)\big)/\bQ_{it}(v_1)\quad k=1,\dots,K_0$$
is the proportion of reads that are assigned to node $v_k$ but cannot be further specified to its child nodes.
This is sum of the $l_1$ distances between two compositional vectors over each branch of the taxonomic tree.

In order to simulate data that mimic real  microbiome count data,  count data on the taxonomic tree are generated based on sampling from  a real 16S microbiome dataset from \cite{GutTongue}, where  the gut (feces), palm and tongue microbial samples of 85 college-age adults were  taken in a range of three months and were  characterized using 16S rRNA  sequencing. Within the gut microbiome samples, counts of reads are summarized on a taxonomic tree that has 1050 nodes from kingdom to species (see Figure~\ref{fig:data_generation}(a)). Since no  large change is expected  in gut microbiome  during a three-month period,  these samples are assumed to have  the same null distribution, which results in a total of  638 gut microbial samples.  Using the notation in Section~\ref{sec:sub_tree}, these samples are denoted as $\bQ^o_1,\dots,\bQ^o_{638}$.  The composition matrix 
 matrix $\textbf{P}^o\in (0,1)^{638\times1050}$ with
 $$\textbf{P}^o(i,k)=\bigg(\bQ^o_{i}(v_k)-\sum_{j:v_j\in\tau(v_k)}\bQ^o_{i}(v_j)\bigg)/\bQ^o_i(v_1), \quad i=1,\dots,638,\quad k=1,\dots,1050$$ is first  calculated, 
which is the composition of all nodes for each of the 638 gut microbial samples.  The total counts of reads of all samples are also calculated  and recorded as  $\textbf{N}^o\in \mathbb{N}^{638}$.

To simulate a pair of correlated microbiome sample $\bQ_{i1}$ and $\bQ_{i2}$,  three compositions $\bP^o_{i1}$, $\bP^o_{i2}$ and $\bP^o_{i3}$ from $\textbf{P}^o$ are randomly sampled and   two total counts $N^o_{i1}$ and $N^o_{i2}$ are randomly resampled from $\textbf{N}^o$. { Read counts  $\mathbf{W}_{i1}$  and $\mathbf{W}_{i2}$  are then  sampled from multinomial distributions $(N^o_{i1}, (\bP^o_{i1}+\bP^o_{i3})/2)+ \mathbf{E}_{i1}$ and $(N^o_{i2}, (\bP^o_{i2}+\bP^o_{i3})/2)+\mathbf{E}_{i2}$,  respectively, where $\mathbf{E}_{i1}$ and $\mathbf{E}_{i2}$ are small perturbations to certain nodes of the tree to generate $\mathbf{W}_{i1}$ and $\mathbf{W}_{i2}$ that have different distributions. Two differential abundance patterns are considered: 
\begin{enumerate}
	\item Sparse differential abundance: $\mathbf{E}_{i1}=0$  and $\mathbf{E}_{i2}$ is drawn from Binomial$(N^o_{i2}, p_{\epsilon})$ at the coordinate corresponding to the genus of {\it Streptococcus} and zero otherwise.
	\item Dense differential abundance: $\mathbf{E}_{i1}$ is drawn from Binomial$(N^o_{i2}, p_{\epsilon})$ at the coordinates corresponding to the genera of {\it Streptococcus, Eubacterium, Parabacteroides}, and zero otherwise, and  $\mathbf{E}_{i2}$ is drawn from Binomial$(N^o_{i2}, p_{\epsilon})$ at the coordinates corresponding to the genera of {\it Porphyromonas, Moraxella, Ruminococcus}, and zero otherwise.
\end{enumerate}	
}
The count vector $\bQ_{it}$ is then iteratively computed such that
$\bQ_{it}(v_k)-\sum_{j:v_j\in\tau(v_k)}\bQ_{it}(v_j)=\mathbf{W}_{itk}$ for $t=1,2$ and $i=1,\dots,n$, where $n$ is the number of pairs simulated and is set to be 20, 50 and 100 in our simulation.
The percent of perturbation $p_{\epsilon}$ is chosen to range from 0 to 2\%. For each scenario,  the simulations are repeated  100 times. For the global test of \eqref{eq:treehypotheses}, the null hypothesis is rejected  at the $\alpha$-level of 0.05. For the identification of subtrees with differential subcompositions in multiple testing \eqref{eq:testing_single}, the FDR is controlled at the  0.05 level.

Figure~\ref{fig:tree_real} compares the rejection rate of PERMANOVA with our method using  \eqref{eq:combined_fisher} or \eqref{eq:combined_2nd} for the global test \eqref{eq:treehypotheses}.  For the method \eqref{eq:combined_2nd} that ombines $p$-values using the $2^{nd}$ smallest $p$-value,  our paired test based on PairMN in \eqref{eq:teststat} is also compared  with the unpaired test based on DM in \eqref{eq:old_test}. { In the sparse differential abundance setting, }when the sample size is small, none of the methods is able to detect  the perturbation to {\it Streptococcus}. As the sample size increases, the rejection rate of our method using the $2^{nd}$ smallest $p$-value combination of the paired-tests gradually increases, especially when the percent of perturbation gets closer to 2\%. Fisher's method combining  $p$-values  does not perform as well because the perturbation only occurs to a very small number of subtrees. The method using $2^{nd}$ smallest $p$-value combination of the unpaired tests also performs worse than the paired tests. 

{ In the dense differential abundance setting, all methods are able to detect the perturbations. The test based on DM has the largest power, but also has inflated type I error. Among the other tests, our test based on PairMN with $2^{nd}$ smallest $p$-value  performs the best with type I error under control.}


\begin{figure}[H]
	\centering
	\begin{subfigure}[b]{1\textwidth}
	\includegraphics[width=1\linewidth]{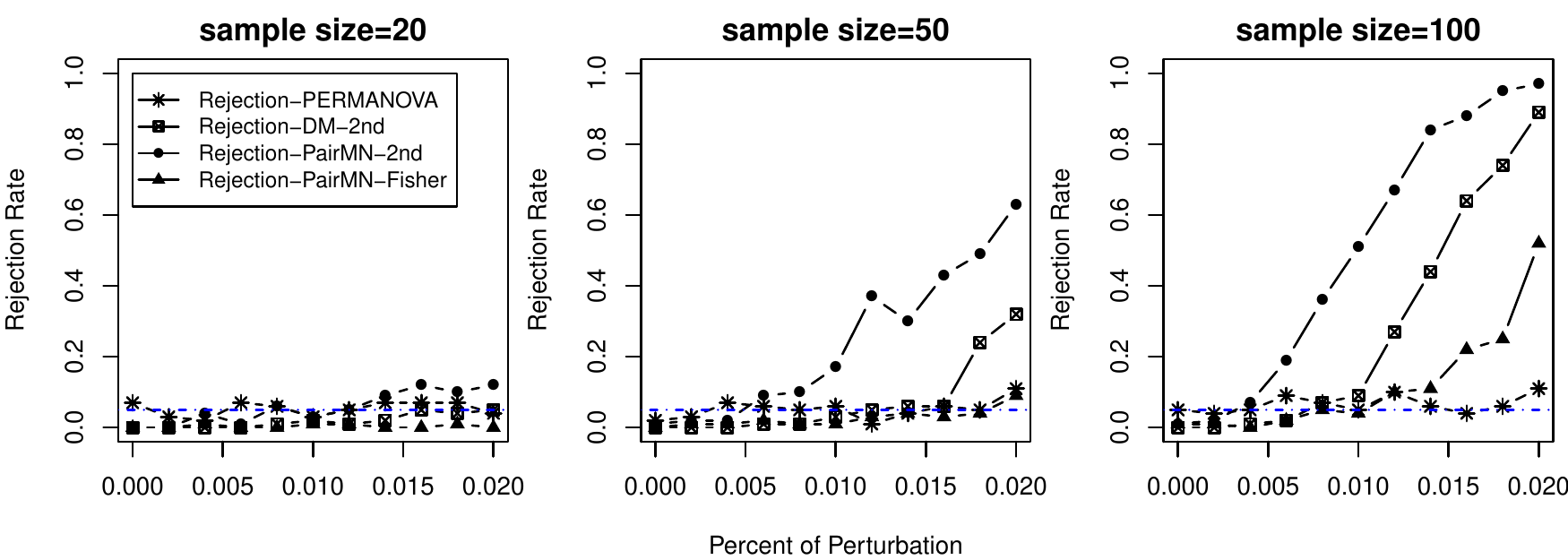}
	\caption{Sparse differential pattern setting.}
	\label{fig:tree_real1}
	\vspace{0.5cm}
	\end{subfigure}
	\begin{subfigure}[b]{1\textwidth}
		\includegraphics[width=1\linewidth]{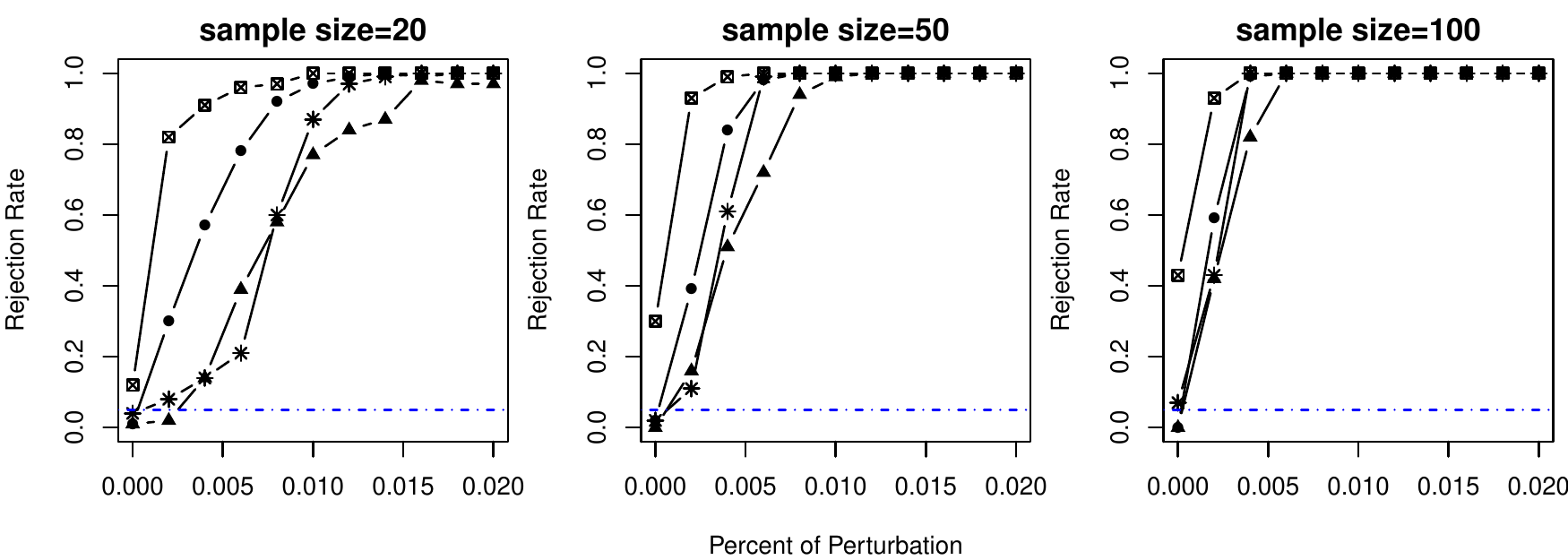}
	\caption{Dense differential pattern setting.}		
		\label{fig:tree_real2}
	\end{subfigure}
\caption{Comparison of rejection rate of the proposed method with PERMANOVA with the level of test at $\alpha=0.05$. x-axis is the perturbation percentage $p_{\epsilon}$, where $p_{\epsilon}=0$ corresponds to the null hypothesis. }\label{fig:tree_real}
\end{figure}

Figure~\ref{fig:real_simu} (a) shows the percent of discoveries of the  differential subtrees { with sparse differential pattern} and FDR controlled at 0.05.  The observed FDR is close to the nominal level of 0.05. Since  the count of genus {\it Streptococcus}  is set to be  different, the counts on all the ancestor nodes of Streptococcus are also changed. Therefore, the differential subtrees denoted by their root nodes are: (a) Kingdom {\it Bacteria}, (b) Phylum {\it Firmicutes}, (c)  Class {\it Bacilli}, (d) Order {\it Lactobacillales}, (e) Family {\it Streptococcaceae} and
(f) Genus {\it Streptococcus} (see Figure~\ref{fig:data_generation} (a)). Among these, (c) and (e) are not identified in any scenario because these subtrees have counts  mostly mapped in one child node and thus make any changes nearly impossible to detect. The test does not have power  to identify (a) because the perturbation is too small to detect given the large counts on the child nodes of (a). All the other three subtrees are identified by our method when the percent of perturbation and sample size get larger.

Figure~\ref{fig:real_simu} (b) shows the percent of discoveries of the  differential subtrees with dense differential abundance and FDR controlled at 0.05. The observed FDR is also close to the nominal level of 0.05. Similar to  the setting of sparse differential abundance, all the ancestor nodes of the perturbed nodes have subcompositions that are different between the two groups. The total number of differential subtrees is 24, but only a subset of these are shown in this figure  since  the other subtrees have undetectable differential subcompositions either because they allocate most counts to one child node, or because the perturbation is too small compared to their base counts.

\begin{figure}[H]
\begin{subfigure}[b]{1\textwidth}
	\centering
	\includegraphics[width=0.85\linewidth]{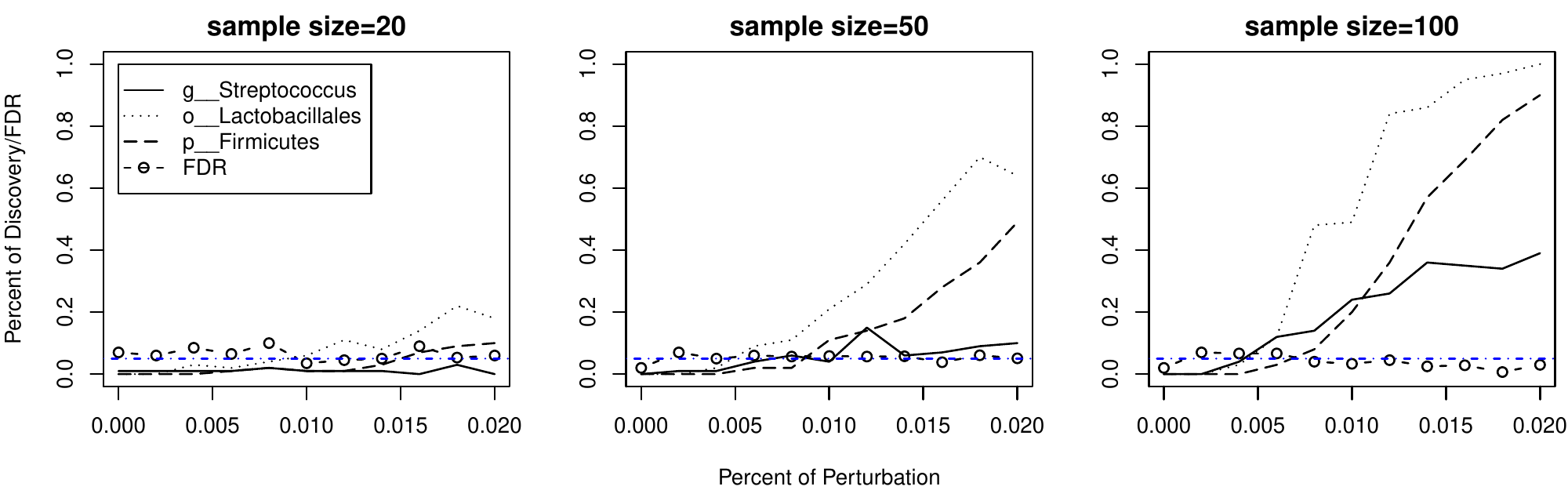}\label{fig:nodes_real}
	\caption{Sparse differential pattern setting.}
	\vspace{0.5cm}
\end{subfigure}
\begin{subfigure}[b]{1\textwidth}
	\centering
	\includegraphics[width=0.85\linewidth]{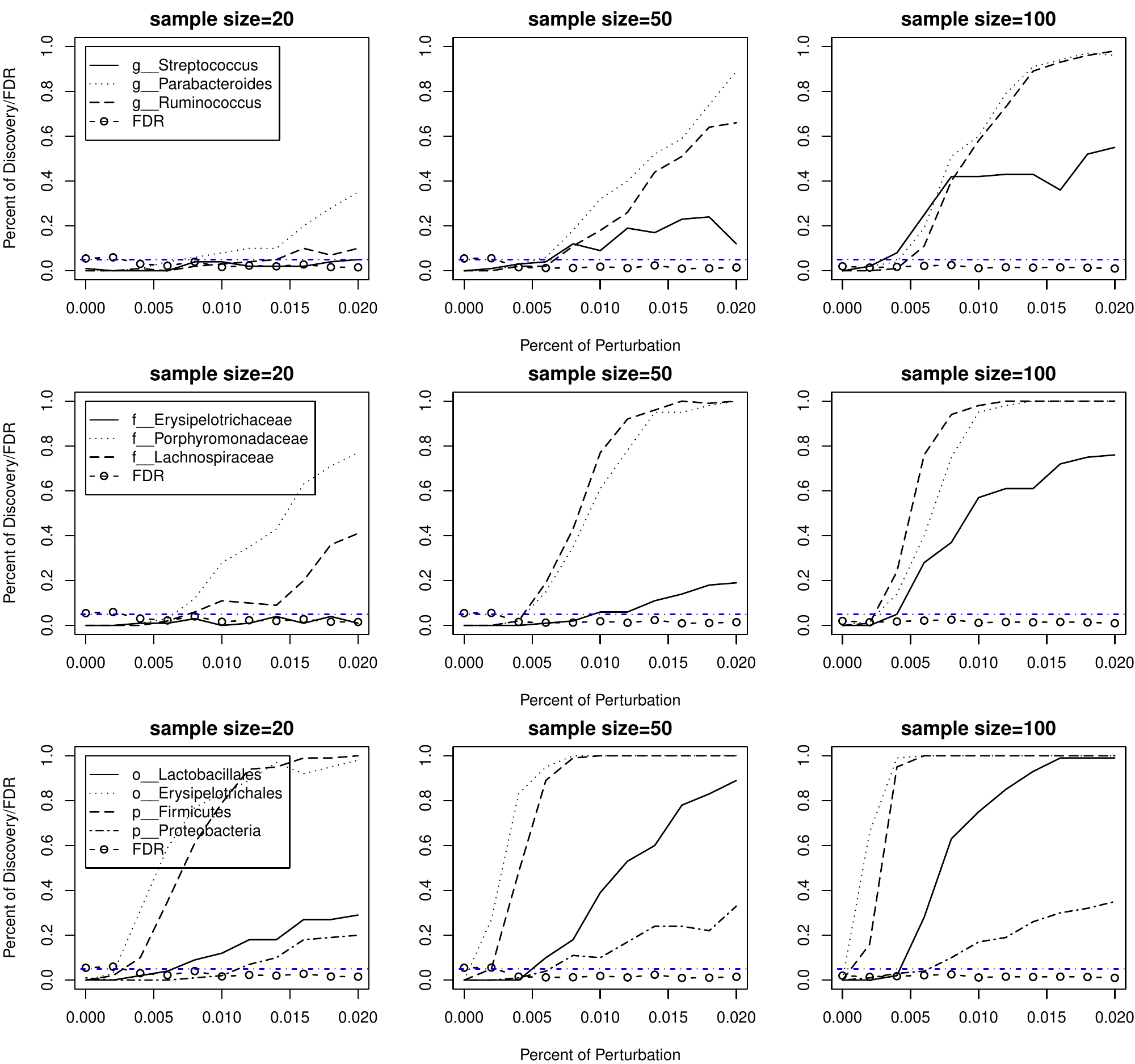}\label{fig:nodes_real2}
	\caption{Dense differential pattern setting.}
\end{subfigure}	
\caption{Identification of subtrees with differential subcompositions with FDR set to 0.05.  y-axis shows the percent of discovery of the corresponding subtree in 100 simulations with FDR controlled at 0.05. The empirical FDR is close to 0.05. In this series of figures,  only  some of the subtrees with parent node that has detectable differential subcompositions are shown. In figure legend, g\_, f\_, o\_, p\_ represent genus, family, order and phylum, respectively. } \label{fig:real_simu}
\end{figure}

\section{Analysis of  Microbiome Data in the Upper Respiratory Tract}\label{sec:real}
The human nasopharynx and oropharynx are two body sites located very close to each other in the upper respiratory tract. The nasopharynx is the ecological niche for many commensal bacteria. 
It is interesting to understand whether these nearby  sites have similar microbiome composition and how  smoking perturbs their compositions.   \cite{SmokerData} collected the left and right nasopharynx and oropharynx microbiome samples from 32 current smokers and 36 nonsmokers. The samples were sequenced using 16S rRNA sequencing, and the count of reads are aligned onto a taxonomic tree with 213 nodes from kingdom to species. { One nonsmoker and one smoker had missing left nasopharynx sample, and one smoker had missing right oropharynx sample.}

Several comparisons of the overall microbiome compositions were compared and the results are summarized in Table \ref{pvalue.tbl}.  As expected, no significant differences were observed between left and right nasopharynx or oropharynx. However very significant differences were observed between nasopharynx and oropharynx both in the left and right sides, further confirming the niche-specific colonization at discrete anatomical sites.  In addition, smoking had strong effects on microbiome composition in both nasopharynx and oropharynx

	\begin{table}
		\caption{$p-$values of different comparisons between two body sites and between smokers and non-smokers based on the proposed tests using the Fisher's method or the 2nd smallest p value and PERMANOVA.}\label{pvalue.tbl}
		\begin{tabular}{lcc}
			\hline
			& PairMN (Fisher $||$ 2nd) & PERMANOVA\\
			\cline{2-3}
		Nasopharynx and Oropharynx  (Left Side)& 0 $||$ 0 & $<$0.001 \\
			Nasopharynx and Oropharynx  (Right Side)& 0 $||$ 0 & $<$0.001 \\
			Smoker vs Nonsmoker (nasopharynx)& 2.1e-07 $||$ 8.6e-05
			& 0.003\\
			Smoker vs Nonsmoker (oropharynx )& 1.2e-07 $||$ 6.2e-04 & 0.005\\
			Left vs Right (nasopharynx)& 0.16 $||$ 0.65 & 0.053\\
			Left  vs Right (oropharynx )& 0.37 $||$ 0.79 & 0.99\\
			\hline
		\end{tabular}
	\end{table}

\subsection{Comparison of nasopharynx and oropharynx microbiome for nonsmokers}
Since a large overall microbiome composition difference was observed, it is interesting to identify which subtrees and their corresponding subcompositions led to such a difference.   The  proposed subtree identification procedure in Section~\ref{sec:testing_tree} using the pairNM  test in \eqref{eq:test} was applied  to identify the subtrees with differential subcompositions between the two body sites at an FDR=0.05.  The identified parental nodes, their child nodes and the corresponding subcompositions are  shown in Figure~\ref{fig:barplot_site} (a).  One advantage of the proposed method is to identify these subtrees at various taxonomic levels.  For example, at the phylum level, nasopharynx clearly had more {\it Firmicutes}, however, oropharynx had more {\it Bacteroidetes}.  At the genus level, {\it Streptococcus} appeared more frequently in oropharynx, but {\it Lactococcus} occurred more in nasopharynx. 


\begin{figure}[h]
\centering
\begin{subfigure}[b]{1\textwidth}
	\centering
	\hspace{1cm}\includegraphics[width=0.60\textwidth]{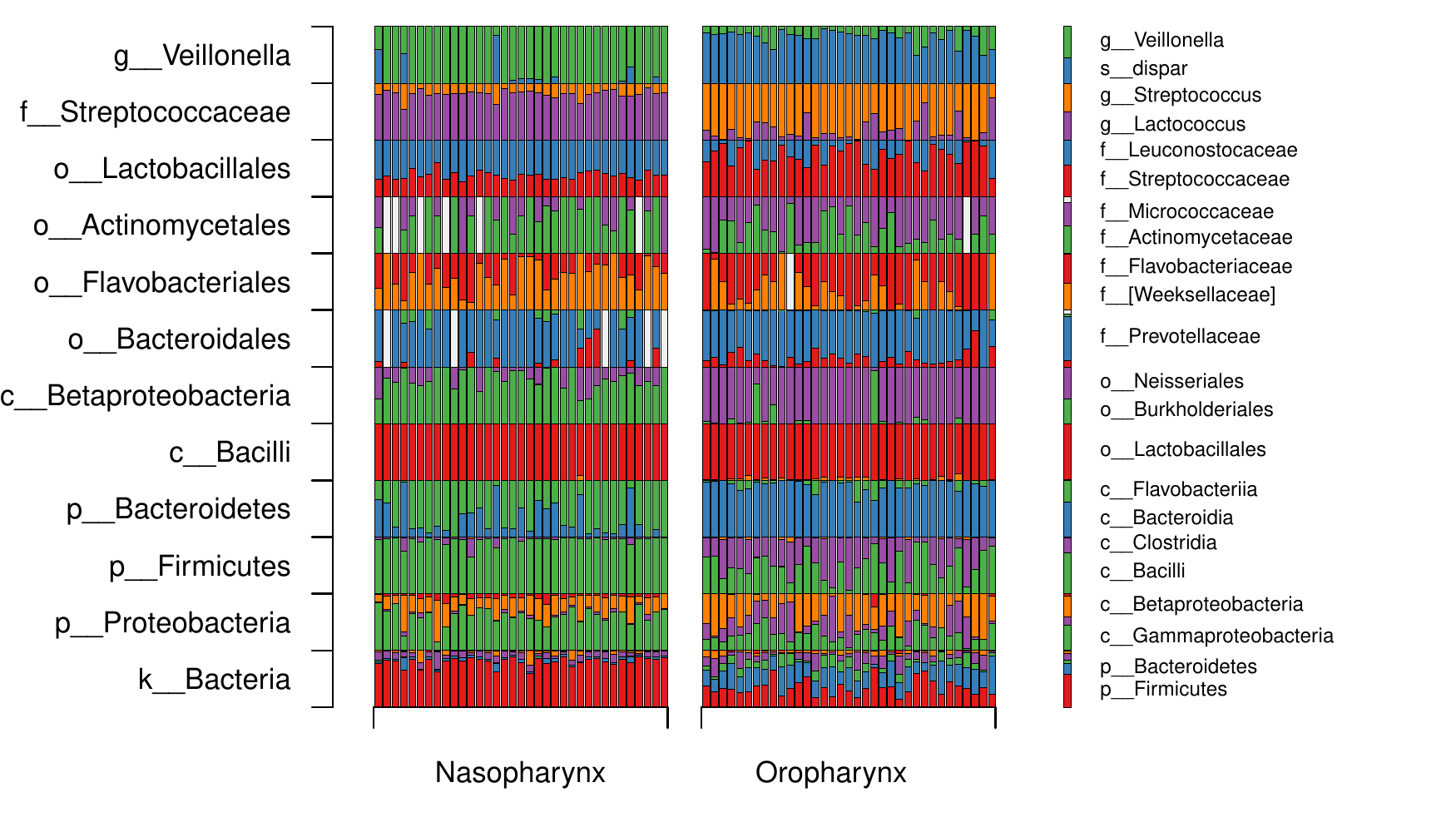}\caption{Comparison between left nasopharynx and left oropharynx among nonsmokers.}
\end{subfigure}

\begin{subfigure}[b]{1\textwidth}
	\centering
	\includegraphics[width=0.56\textwidth]{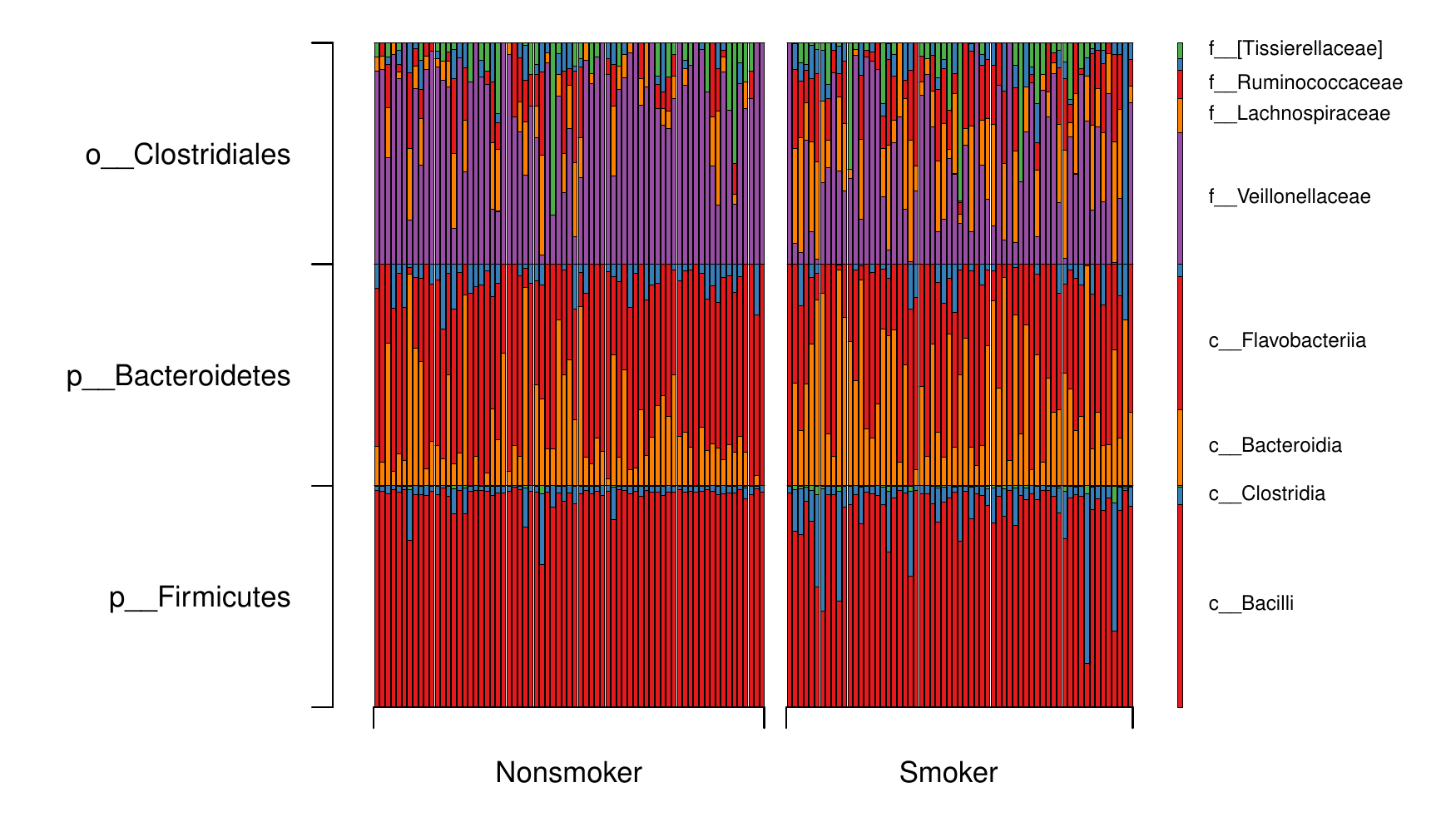}\caption{Comparison between smokers and nonsmokers in nasopharynx.}
\end{subfigure}

\begin{subfigure}[b]{1\textwidth}
	\centering
	\includegraphics[width=0.56\textwidth]{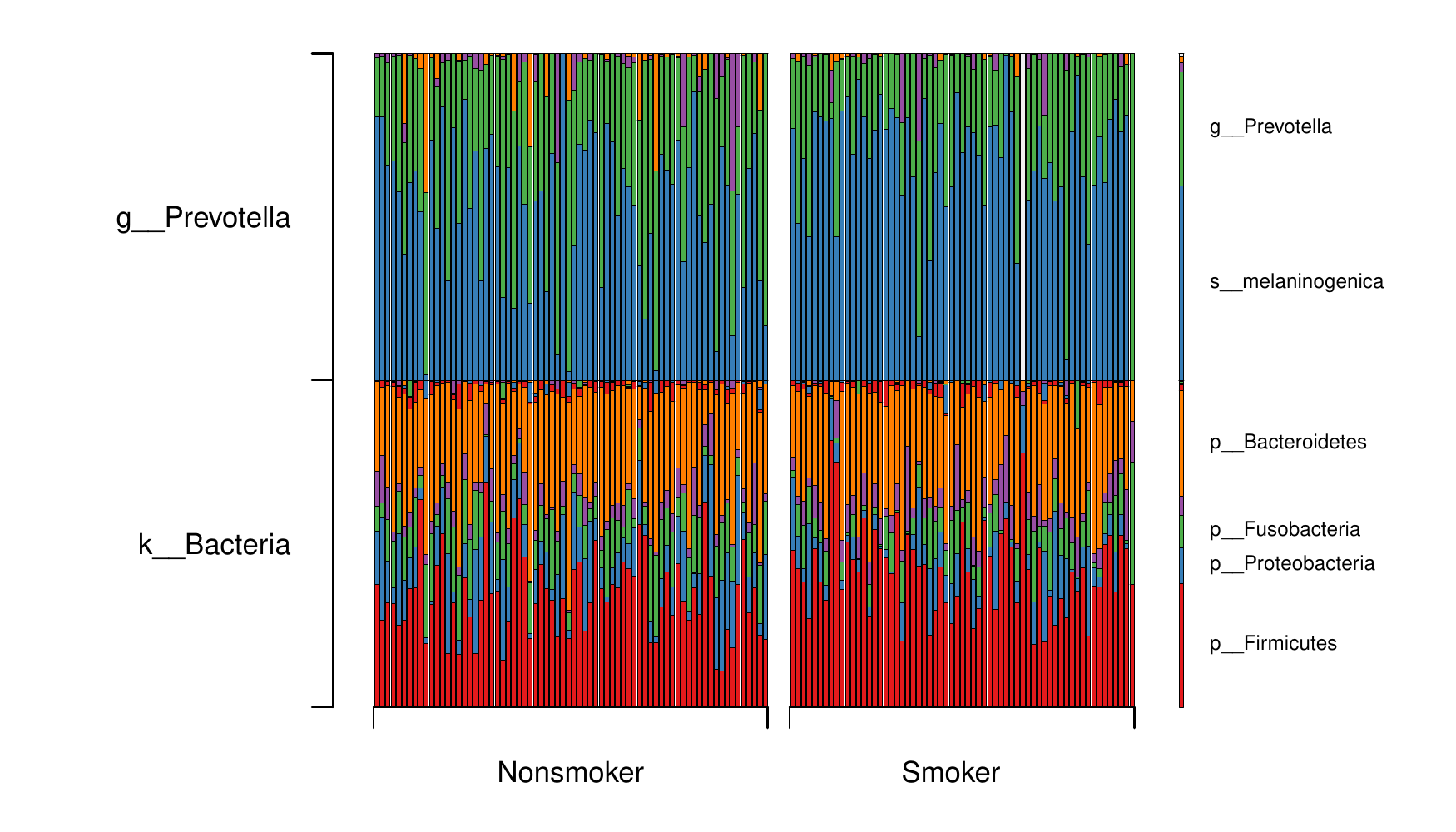}\caption{Comparison between smokers and nonsmokers in oropharynx.}
\end{subfigure}

\caption{Analysis of upper respiratory tract microbiome data. Parental nodes and the child nodes that showed differential subcomposition are presented. In taxon labels, g\_, f\_, o\_, p\_ represent genus, family, order and phylum, respectively. For (a), the sample size is 35. The number of parameters within each subtree ranges from 2 to 6. For (b), the sample size is 71 for nonsmokers and 63 for smokers. The number of parameters to test within each subtree ranges from 2 to 6. For (c), the sample size is 72 for nonsmokers and 63 for smokers. The number of parameters to test within each subtree ranges from 2 to 10.}
\label{fig:barplot_site}
\end{figure}

\subsection{Comparison of microbiome between smokers and non-smokers}
The proposed procedure was also applied to identify the differential subtrees with differential subcompositions between smokers and nonsmokers in nasopharynx and oropharynx and the results are shown in Figures \ref{fig:barplot_site} (b) and (c) for an FDR=0.05.  For nasopharynx, the subcompositions of classes under Phylum {\it Firmicutes}, classes under Phylum {\it Bacteroidetes} and families under Order {\it Clostridiale} were different, with fewer {\it Bacilli} in {\it Firmicutes}, more {\it Bacteroidia} in {\it Bacteroidetes}, and fewer  {\it Veillonellaceae} in {\it Clostridiales} being observed in smokers (Figure~\ref{fig:barplot_site} (b)). 

For oropharynx,  differences in the subcomposition of phyla and species under Genus {\it Prevotalla} were observed,  with more {\it Firmicuates} in Kingdom {\it Bacteria} and more {\it Melaninogenica} in Genus {\it Prevotella} observed in smokers (Figure~\ref{fig:barplot_site} (c)).

\section{Discussion}\label{sec:discuss}
This paper has introduced a flexible model for paired multinomial data. Based on this model,  a $T^2$-type of test statistic has been developed  for testing equality of the overall taxa composition between two repeatedly measured multinomial data.  The test can be  used  for analysis of count data observed on a taxonomic tree to identify the subtrees that show differential subcompositions in repeated measures. Our simulations have shown that the proposed test has correct type 1 errors and much increased power than the commonly used tests based on DM model or the PERMANOVA test.   The test proposed in this paper can be applied to both independent and repeated measurement data.  For independent data, the proposed test allows more flexible dependency structure among the taxa than the Dirichlet-multinomial model, which only allows negative correlations among the taxa.  The proposed test statistics are also computationally more efficient than the commonly used permutation-based procedures such as PERMANOVA, which enables their applications in large-scale microbiome studies.

As demonstrated in our simulations, the proposed overall test of composition is more powerful than PERMANOVA type of tests when the overall composition difference is due to a few subcompositions since our test considers each subtree and subcomposition separately and then combines the $p$-values. Since the tests for differential subcomposition condition on the total counts of the parental nodes, all the $p$-values are independent, which facilitates simple combination of $p$-values and identification of the subtrees based on FDR controlling. In general,  test based on the $2^{nd}$ smallest $p$-value for overall composition difference  is more sensitive than the Fisher's method, as is shown in simulations in Section \ref{sec:simu_tree}. However, the Fisher's method is expected to have higher power when most of the subtrees show differential compositions. When the differential pattern is not known, one possible solution is to take the smallest of these two $p$-values and to use permutations to assess its significance. 

Although the paper has focused on using existing taxonomic tree and 16S sequencing data, the tests proposed in this paper can also be applied to shotgun metagenomic sequencing data. One possible approach is to build phylogenetic trees based on a small set of universal marker genes \citep{UniMarker} and to align the sequencing reads to these phylogenetic trees.  The proposed methods can be applied to each of these trees and the results can be combined. This deserves further investigation.

\label{lastpage}

\newpage
\section*{Appendix - Proofs}

\textbf{Proof of Lemma~\ref{th:test}}

By \eqref{eq:moments}, we have
\begin{eqnarray}
\Var \hat{\bpi}_t&=&\dfrac{\sum_{i=1}^nN_{it}^2-N_{\cdot t}}{N_{\cdot t}^2}\bSigma_t + \dfrac{1}{N_{\cdot t}}\big(diag(\bpi_t)-\bpi_t\bpi_t^\top\big),\nonumber\\
\Cov(\hat{\bpi}_1, \hat{\bpi}_2)&=&\dfrac{\sum_{i=1}^n N_{i1}N_{i2}}{N_{\cdot 1}N_{\cdot 2}}\bSigma_{12}.\nonumber
\end{eqnarray}
It can also be shown that
\begin{eqnarray}
\Exp(\bS_t-\bG_t)&=& N_{ct}\bSigma_t,\nonumber\\
\Exp\big(\bS_t + (N_{ct}-1)\bG_t\big)&=& N_{ct}\big(diag(\bpi_t)-\bpi_t\bpi_t^\top\big),\nonumber\\
\Exp \widehat{\bSigma}_{12}&=& \bSigma_{12}. \nonumber
\end{eqnarray}
Thus,
\begin{equation}\label{eq:sigma}
\begin{split}
\Sigpi~=~&\Var(\hat{\bpi}_1-\hat{\bpi}_2)\\
~=~&\sum_{t=1}^2\left\{\dfrac{\sum_{i=1}^nN_{it}^2-N_{\cdot t}}{N_{\cdot t}^2}\bSigma_t+\dfrac{1}{N_{\cdot t}}\big(diag(\bpi_t)-\bpi_t\bpi_t^\top\big)\right\}-\dfrac{2\sum_{i=1}^n N_{i1}N_{i2}}{N_{\cdot 1}N_{\cdot 2}}\bSigma_{12}\\
~=~&\sum_{t=1}^2\left\{\dfrac{\sum_{i=1}^nN_{it}^2-N_{\cdot t}}{N_{ct}N_{\cdot t}^2}\Exp(\bS_t-\bG_t)+\dfrac{1}{N_{ct}N_{\cdot t}}\Exp\big(\bS_t + (N_{ct}-1)\bG_t\big)\right\}\\
&~-~\dfrac{\sum_{i=1}^n N_{i1}N_{i2}}{N_{\cdot 1}N_{\cdot 2}}\Exp \big(\widehat{\bSigma}_{12}+\widehat{\bSigma}_{12}^\top\big).
\end{split}
\end{equation}
{By law of large numbers, the following convergences hold  in probability as $n\rightarrow\infty$.}
\begin{equation}\label{eq:CLT}
\begin{split}
||(\bS_t-\bG_t)-\Exp(\bS_t-\bG_t)||_{\max}~\rightarrow~& 0,\\
||\big(\bS_t + (N_{ct}-1)\bG_t\big)-\Exp\big(\bS_t + (N_{ct}-1)\bG_t\big)||_{\max}~\rightarrow~& 0,\\
||\bSigma_{12}-\Exp \widehat{\bSigma}_{12}||_{\max}~\rightarrow~& 0.
\end{split}
\end{equation}
Combining \eqref{eq:sigmahat}, \eqref{eq:sigma} and \eqref{eq:CLT}, we have
\begin{eqnarray}
||\Sighatpi - \Sigpi||_{\max} \rightarrow 0\quad \mbox{ in probability as }n\rightarrow\infty\nonumber
\end{eqnarray}

\newpage
\textbf{Proof of Theorem~\ref{th:test}}

Define $S^{d-1}=\{x\in\mathbb{R}^p:\boldsymbol{1}^\top x=0\}$. Then $\bpi_1-\bpi_2,\hat{\bpi}_1-\hat{\bpi}_2\in S^{d-1}$. Therefore
$$(\hat{\bpi}_1-\hat{\bpi}_2)^\top\Sigpi^{\dagger}(\hat{\bpi}_1-\hat{\bpi}_2)\rightarrow \chi_{d-1}^2.$$
We   show next that $\Sighatpi^\dagger\rightarrow\Sigpi^\dagger$ in probability.  Then by Slutsky Theorem, for fixed $d$ and $n\rightarrow\infty$,
$$F=\dfrac{n-d+1}{(n-1)(d-1)}(\hat{\bpi}_1-\hat{\bpi}_2)^\top\Sighatpi^{\dagger}(\hat{\bpi}_1-\hat{\bpi}_2)\rightarrow \chi_{d-1}^2/(d-1)$$
Since $F_{d-1, n-d+1}\rightarrow \chi^2_{d-1}/(d-1)$ for fixed $d$ and $n\rightarrow\infty$, $F>F_{d-1, n-d+1}^{-1}(1-\alpha)$ is an asymptotic level $\alpha$ test.

Let $\boldsymbol{\Gamma}$ be an orthogonal matrix in the form of $[\mathbf{V},\boldsymbol{1}^\top_d/\sqrt{d}]$. Then, because $\boldsymbol{1}_d^\top \Sighatpi=\boldsymbol{1}_d^\top\Sigpi=0$, by Lemma~\ref{th:covariance}, we have
\begin{equation*}
\begin{split}
||\mathbf{V}^\top (\Sighatpi-\Sigpi)\mathbf{V}||_2/d \leq ||\mathbf{V}^\top (\Sighatpi-\Sigpi)\mathbf{V}||_{\max}=||\mathbf{\Gamma}^\top (\Sighatpi-\Sigpi)\mathbf{\Gamma}||_{\max}\rightarrow 0 \quad \mbox{in probability}
\end{split}
\end{equation*}
where $||\cdot||_2$ is the spectral norm of matrix. 
Define $\boldsymbol{\Delta} = \mathbf{V}^\top (\Sighatpi-\Sigpi)\mathbf{V}$.
Using Neumann series expansion,
\begin{equation*}
\begin{split}
(\mathbf{V}^\top \Sighatpi \mathbf{V})^{-1} - (\mathbf{V}^\top\Sigpi \mathbf{V})^{-1}&=(\mathbf{V}^\top\Sigpi \mathbf{V})^{-1}\sum_{i=1}^\infty \big(-(\mathbf{V}^\top\Sigpi \mathbf{V})\boldsymbol{\Delta}\big)^i,
\end{split}
\end{equation*}
therefore, 
$$||(\mathbf{V}^\top \Sighatpi \mathbf{V})^{-1} - (\mathbf{V}^\top\Sigpi \mathbf{V})^{-1}||_2\leq \sum_{i=0}^\infty ||\mathbf{V}^\top\Sigpi \mathbf{V}||_2^i||\boldsymbol{\Delta}||_2^{i+1}.
$$
{Because $||\mathbf{V}^\top\Sigpi\mathbf{V}||_2$ is fixed, we have $||\mathbf{V}^\top\Sigpi\mathbf{V}||_2||\boldsymbol{\Delta}||_2\rightarrow 0$ in probability, which implies $\Prob\big(||\mathbf{V}^\top\Sigpi\mathbf{V}||_2||\boldsymbol{\Delta}||_2<1\big)\rightarrow 1.$
Therefore, with probability one, $$\sum_{i=0}^\infty ||\mathbf{V}^\top\Sigpi \mathbf{V}||_2^i||\boldsymbol{\Delta}||_2^{i+1}
=\dfrac{||\boldsymbol{\Delta}||_2}{1-||\mathbf{V}^\top\Sigpi \mathbf{V}||_2||\boldsymbol{\Delta}||_2}$$
holds.  As a result,
\begin{equation*}
\begin{split}
||(\mathbf{V}^\top \Sighatpi \mathbf{V})^{-1} - (\mathbf{V}^\top\Sigpi \mathbf{V})^{-1}||_{\max} &\leq ||(\mathbf{V}^\top \Sighatpi \mathbf{V})^{-1} - (\mathbf{V}^\top\Sigpi \mathbf{V})^{-1}||_2 \nonumber\\
&\leq \dfrac{||\boldsymbol{\Delta}||_2}{1-||\mathbf{V}^\top\Sigpi \mathbf{V}||_2||\boldsymbol{\Delta}||_2} \rightarrow 0 \quad \mbox{in probability}, \nonumber 
\end{split}
\end{equation*}}
which leads to 
\begin{equation}\label{eq:VV}
||\mathbf{V}(\mathbf{V}^\top \Sighatpi \mathbf{V})^{-1}\mathbf{V}^\top - \mathbf{V}(\mathbf{V}^\top\Sigpi \mathbf{V})^{-1}\mathbf{V}^\top||_{\max}\rightarrow 0 \quad \mbox{in probability}.
\end{equation}

Suppose we have the eigenvalue decomposition of $\Sigpi$ as $\Sigpi=\mathbf{U}\boldsymbol{\Lambda} \mathbf{U}^\top$, where $\mathbf{U}\in\mathbb{R}^{d\times(d-1)}$ and $\boldsymbol{\Lambda}\in\mathbb{R}^{(d-1)\times(d-1)}$. 
Then $\boldsymbol{1}_d^\top \mathbf{U}=0$. Also, $\mathbf{U}^\top\mathbf{V}$ is orthogonal because
$$\mathbf{U}^\top\mathbf{V}\mathbf{V}^\top \mathbf{U}=\mathbf{U}^\top(\textbf{I}_d-\boldsymbol{1}_d\boldsymbol{1}_d^\top /d)\mathbf{U} = \mathbf{U}^\top\mathbf{U} = \textbf{I}_{d-1}$$
Therefore,
\begin{equation*}
\begin{split}
& \mathbf{V} (\mathbf{V}^\top\Sigpi \mathbf{V})^{-1} \mathbf{V}^\top= \mathbf{V} (\mathbf{V}^\top \mathbf{U}\boldsymbol{\Lambda} \mathbf{U}^\top \mathbf{V})^{-1} \mathbf{V}^\top\\
&= \mathbf{V}(\mathbf{U}^\top \mathbf{V})^{-1}\boldsymbol{\Lambda}^{-1}(\mathbf{V}^\top \mathbf{U})^{-1}\mathbf{V}^\top = \mathbf{V}(\mathbf{U}^\top \mathbf{V})^\top\boldsymbol{\Lambda}^{-1}(\mathbf{V}^\top \mathbf{U})^\top \mathbf{V}^\top\\
&= (\textbf{I}_d-\boldsymbol{1}_d\boldsymbol{1}_d^\top /d) \mathbf{U}\boldsymbol{\Lambda}^{-1}\mathbf{U}^\top (\textbf{I}_d-\boldsymbol{1}_d\boldsymbol{1}_d^\top /d)= \mathbf{U}\boldsymbol{\Lambda}^{-1}\mathbf{U}^\top\\
&=\Sigpi^{\dagger}.
\end{split}
\end{equation*}

Similarly, we have
$$\Sighatpi^\dagger=\mathbf{V} (\mathbf{V}^\top\Sighatpi \mathbf{V})^{-1} \mathbf{V}^\top.$$
The proof of this statement  is similar to the proof of $\Sigpi^\dagger=\mathbf{V} (\mathbf{V}^\top\Sigpi \mathbf{V})^{-1} \mathbf{V}^\top$. Because $n>d$, we still have the eigenvalue decomposition of $\Sighatpi$ as $\Sighatpi=\mathbf{U}\boldsymbol{\Lambda} \mathbf{U}^\top$, where $\mathbf{U}\in\mathbb{R}^{d\times(d-1)}$ and $\boldsymbol{\Lambda}\in\mathbb{R}^{(d-1)\times(d-1)}$. 
Then $\boldsymbol{1}_d^\top \mathbf{U}=0$. Also, $\mathbf{U}^\top\mathbf{V}\in\mathbb{R}^{(d-1)\times(d-1)}$ is orthogonal because
$\mathbf{U}^\top\mathbf{V}\mathbf{V}^\top \mathbf{U}=\mathbf{U}^\top(\textbf{I}_d-\boldsymbol{1}_d\boldsymbol{1}_d^\top /d)\mathbf{U} = \mathbf{U}^\top\mathbf{U} = \textbf{I}_{d-1}.$

Further, we show that $\Sighatpi^\dagger$ is indeed a generalized inverse of $\Sighatpi$, because 
\begin{equation*}
\begin{split}
\Sighatpi^\dagger\Sighatpi\Sighatpi^\dagger
&=\bV(\bV^\top\bU\bLambda\bU^\top\bV)^{-1}\bV^\top \bU\bLambda\bU^\top\bV(\bV^\top\bU\bLambda\bU^\top\bV)^{-1}\bV^\top \\
&=\bV(\bU^\top\bV)^{-1}\bLambda^{-1}(\bV^\top\bU)^{-1}\bV^\top \bU\bLambda\bU^\top\bV(\bU^\top\bV)^{-1}\bLambda^{-1}(\bV^\top\bU)^{-1}\bV^\top\\
&=\bV(\bU^\top\bV)^{-1}\bLambda^{-1}(\bV^\top\bU)^{-1}\bV^\top\\
&=\bV(\bU^\top\bV)^\top\bLambda^{-1}(\bV^\top\bU)^\top\bV^\top=(\bV\bV^\top)\bU\bLambda^{-1}\bU^\top(\bV\bV^\top)\\
&=(\textbf{I}_{d}-\textbf{1}_d\textbf{1}_d/d)\bU\bLambda^{-1}\bU^\top(\textbf{I}_{d}-\textbf{1}_d\textbf{1}_d/d)\\
&=\bU\bLambda^{-1}\bU^\top=\Sighatpi^\dagger.
\end{split}
\end{equation*}

Based on  \eqref{eq:VV}, we have
$$||\Sighatpi^{\dagger}-\Sigpi^{\dagger}||_{\max} \rightarrow 0 \mbox{ in probability}.$$

\bibliographystyle{apa}
\bibliography{paired_bib}

\end{document}